\documentclass[aps,prl,twocolumn,showpacs,superscriptaddress,groupedaddress]{revtex4}  
\usepackage{epsfig}
\usepackage{graphicx}
\usepackage{subfigure}
\usepackage{graphicx}  
\usepackage{dcolumn}   
\usepackage{bm}        
\usepackage{amssymb}   
\begin{document}


\author{Somayeh Farhadi and Robert P. Behringer}
\affiliation{Department of Physics and Center for Nonlinear
and Complex Systems, Box 90305,
Duke University, Durham, NC 27708}
\date{\today}

\title{Dynamics of Sheared Ellipses and Circular Disks: Effects of
  Particle Shape}

\renewcommand{\textfraction}{0.05}
\renewcommand{\topfraction}{0.95}
\renewcommand{\bottomfraction}{0.95}
\setcounter{bottomnumber}{4} 
\setcounter{topnumber}{4}
\renewcommand{\floatpagefraction}{0.95}

\newcommand{\ea}{{\it et al.}}

\begin{abstract}        

Much recent effort has focused on glassy and jamming properties
of spherical particles.  Very little is known about such phenomena for
non-spherical particles, and we take a first step by studying
ellipses.  We find important differences between the
dynamical and structural properties of disks and two-dimensional
ellipses subject to continuous Couette shear.  In particular, ellipses
show slow dynamical evolution, without a counterpart in disks, in the
mean velocity, local density, orientational order, and local stress.
Starting from an unjammed state, ellipses can first jam under shear,
and then slowly unjam. The slow unjamming process is understood as a
result of gradual changes in their orientations, leading to a denser
packing. For disks, the rotation of particles only contributes to
relaxation of frictional forces, and hence, does not significantly
cause structural changes. For the shear-jammed states, the global
building up and relaxation of stress, which occurs in the form of
stress avalanches, is qualitatively different for disks and ellipses,
and is manifested by different forms of rate-dependence for ellipses
vs. disks.  Unlike the weak rate dependence typical for many granular
systems, ellipses show power-law dependence on the shearing rate,
$\Omega$.

\end{abstract}

\maketitle

There has been recent interest in glassy dynamics,
jamming\cite{ohern03,vanhecke10,bi11} and
rheology\cite{heussinger,teitel,howell99,miller96,hartley03,behringer08,dijksman11}
of particulate systems, such as colloids, emulsions, foams and
granular materials.  Despite variations in interaction details,
particulate systems show similar rheology \cite{Schall}.
Traditionally, rheology concerns aging, rate effects, shear
band formation, and failure
\cite{miller96,geng03,marone,hartley03,geng05,behringer08,dijksman11,Schall}.
And, steadily sheared systems provide an alternative route for
approaching the jamming transition for particulate systems by moving
along the yield stress curve which separates static (jammed) and
flowing (unjammed) states in a parameter space of shear stress,
$\tau$, and packing fraction, $\phi$.

Many recent rheology studies involved spherically symmetric particles.
The question we pose is, how does breaking of spherical particle symmetry change the rheology of near-jamming granular materials.  A
complete answer to this question is beyond the scope of a single
paper.  Here, we take a first step towards understanding the role
played by particle shape by considering elliptical particles. We show
that steadily sheared elliptical particles show dramatically stronger
rate dependence than their circular counterparts.  This rate
dependence appears to be linked to slow evolution of the particle
orientation.  Although we have focused on ellipses because they are
easy to characterize, there are key aspects that ellipses share with
more complex particle shapes: particle orientation and rotation couple
to density and normal contact forces between particles lead to
torques.

The ability of particles to exert torques on each other impacts
jamming in important ways.  For the torque-free case of frictionless
spheres, previous simulations\cite{ohern03} reported a fraction,
$\phi_J$, below which the system is always stress-free (both the
pressure, $P$ and $\tau$ vanish), and the shear modulus vanishes.
Above $\phi_J$, there exist $\tau = 0$, $P >0$ mechanically stable
states.  In the frictionless spherical case, if $\phi < \phi_J$,
$\dot{\gamma}$ must be non-zero to obtain non-zero $\tau$.  However,
Bi et al.\cite{bi11} recently showed that systems of frictional disks
behave differently.  There exists a range $\phi_S \leq \phi \leq
\phi_J$ where it is possible to shear jam a system at fixed $\phi$ by
applying shear strain, starting from a $\tau = P = 0$ state, and
arriving at a jammed (mechanically stable) state.  Here, $\phi_J$
corresponds to the lowest density at which frictional granular systems
can be jammed at zero $\tau$, and $\phi_S$ corresponds to the lowest
density at which it is possible to jam a system by applying shear
strain.  These observations are consistent with simulations along the
yield stress curve by Otsuki and Hayakawa\cite{otsuki}, and older
experiments by Howell et al.\cite{howell99} who observed a what is now
understood as a lower limit for shear jamming $\phi_S \simeq 0.76$ for
bi-disperse frictional disks subject to Couette shear.  Thus,
quasi-static shear of frictional spheres (disks in 2D) in the range
$\phi_S \leq \phi \leq \phi_J$ is clearly affected by the possibility
of shear jamming due to the presence of rotational degrees of freedom.
However, for non-spherical particles, such as ellipses, rotational
modes are imposed by the geometry even for the frictionless case This
experimental study directly probes the effect of particle asphericity
on shear jamming of particulate systems.
 
A key issue for rheology is the dependence of the shear stress and
other properties of a shear system in the strain rate, $\dot{\gamma}$.
Here, we contrast two particular cases: a Newtonian fluid has $\tau
\propto \dot{\gamma}$, and an ideal granular material has $\tau$
independent of $\dot{\gamma}$ for very slow $\dot{\gamma}$ \cite{Savage}.
Although rate-independence for stresses in slowly sheared granular
materials has frequently been assumed, there are multiple granular
experiments\cite{marone,behringer08,reddy11} showing logarithmic rate
dependence in $\dot{\gamma}$.  The true origin of this rate dependence
is still an open question, although several groups have proposed
models in the spirit of soft glassy
rheology\cite{behringer08,reddy11}.  However, as with many (although
not all\cite{donev04,mailman09}) studies of jamming, previous work on
rate dependence, has typically involved systems of spherical or
circular particles, in 3D or 2D, respectively.  Here, we show that
elliptical particles under sustained shear show stronger rate
dependence than their circular counterparts.  By inference, this rate
dependence is linked to particle orientation.

Here, we directly compare sheared systems of ellipses and of
bidisperse disks, and find several crucial differences: 1) Very long
transients for $P$ of ellipses but not for disks; there is a packing
fraction range for ellipses where $P$ builds up relatively quickly,
and then relaxes very slowly. For disks, only monotonic evolution
occurrs. 2) The order parameter, the density, and the velocity profile
exhibit very slow evolution for the ellipse systems; 3) The power
spectrum of stress fluctuations for ellipses suggests power-law
scaling with rate, with an exponent of $\sim 1/2$; spectra for disks
have much weaker rate-dependence.

\begin{figure}
\subfigure[]{\includegraphics[scale=0.16]{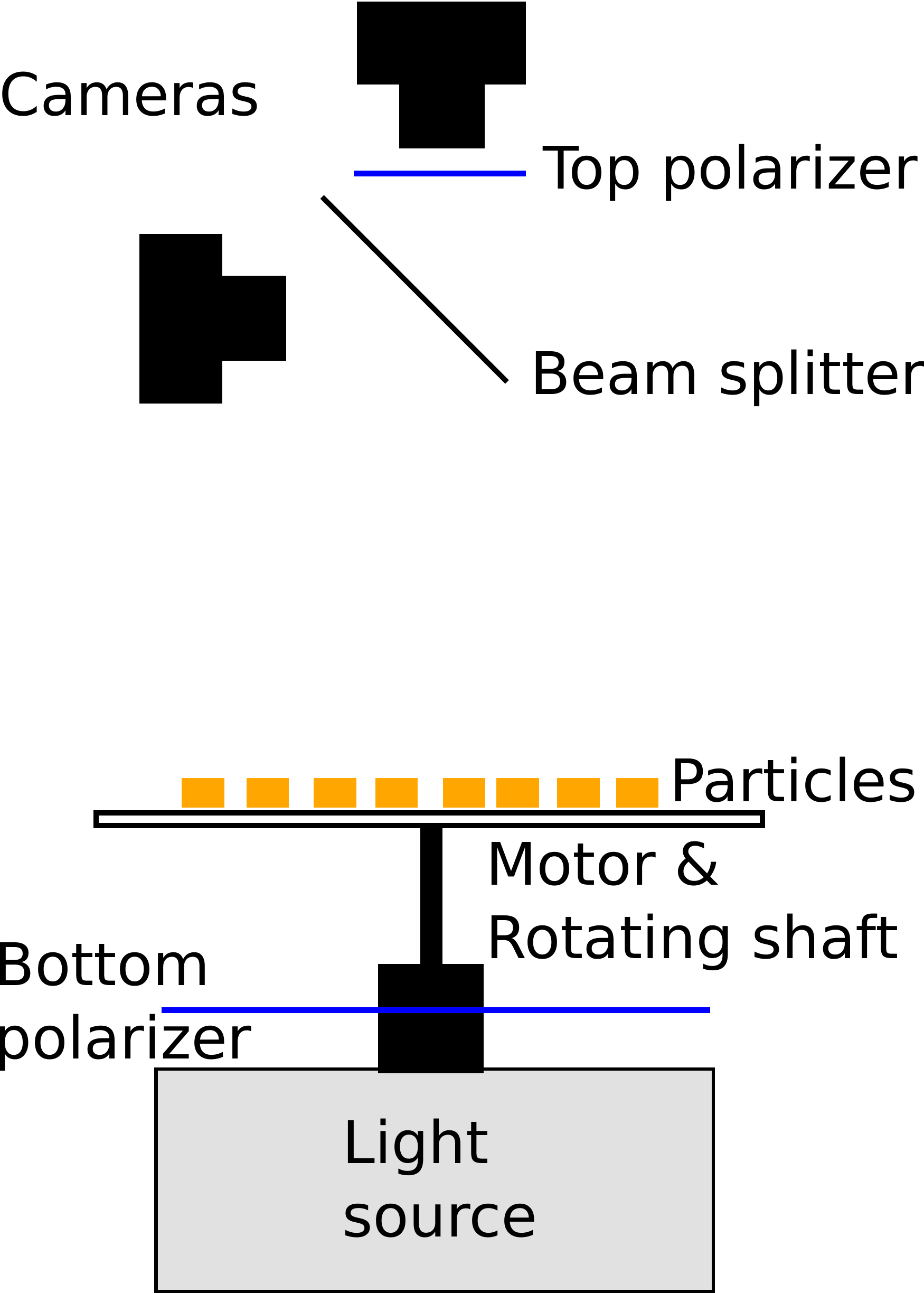}}
\hspace{0.8 cm}
\subfigure[]{\includegraphics[scale=0.21]{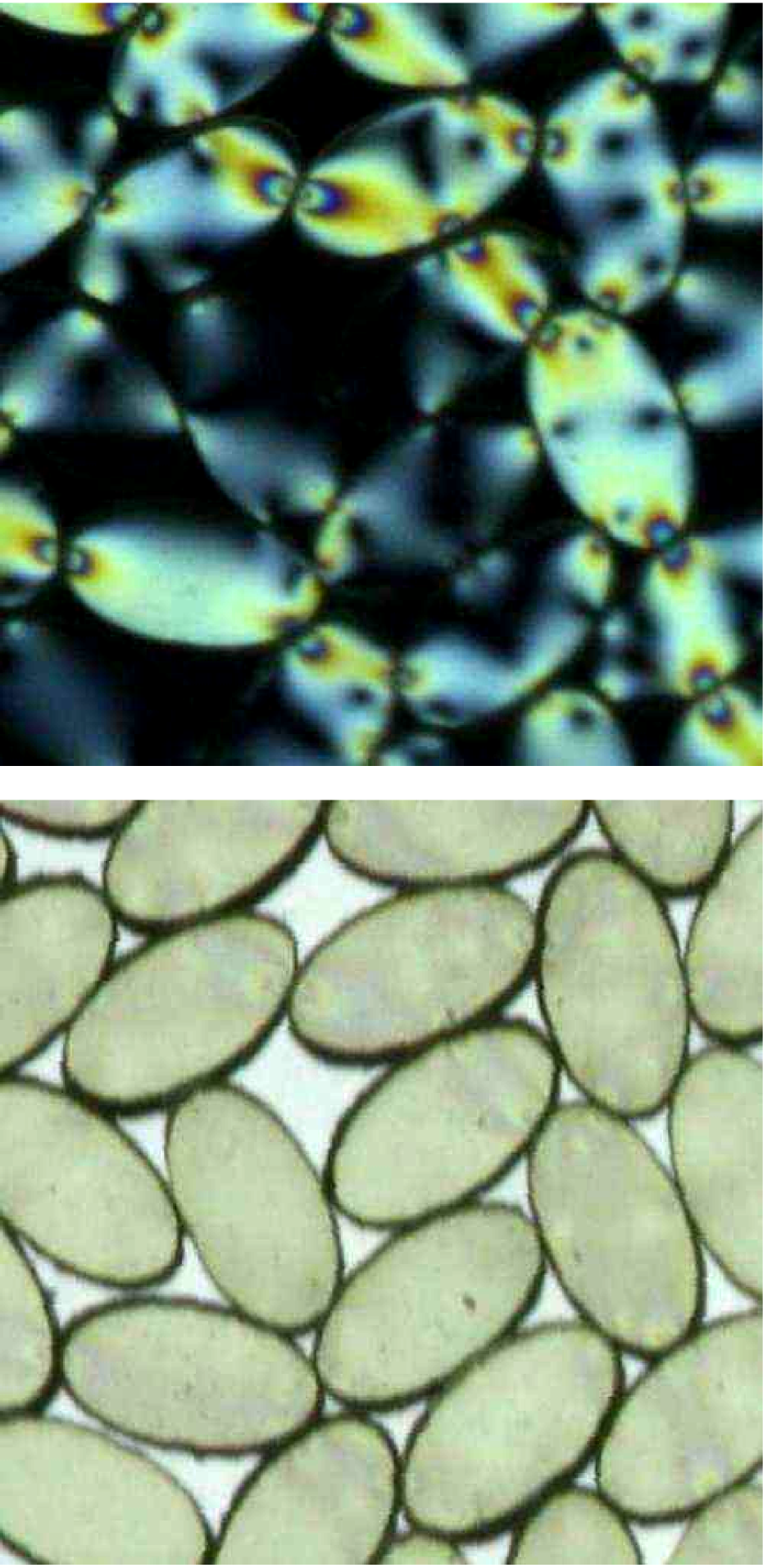}}\\
\subfigure[]{\includegraphics[scale=0.13]{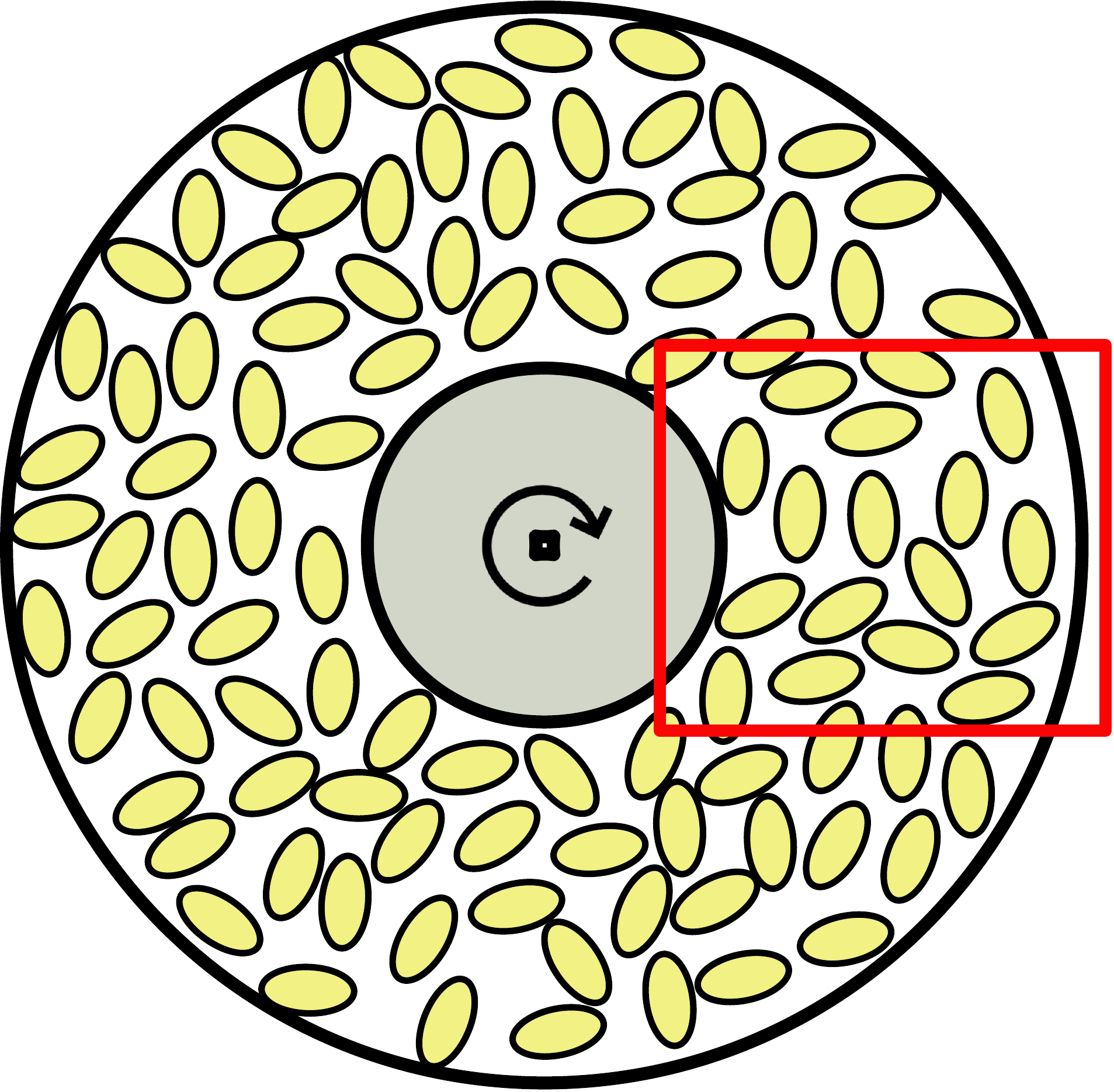}}
\hspace{.1cm}
\hspace{0.2 cm}
\subfigure[]{\includegraphics[scale=0.3]{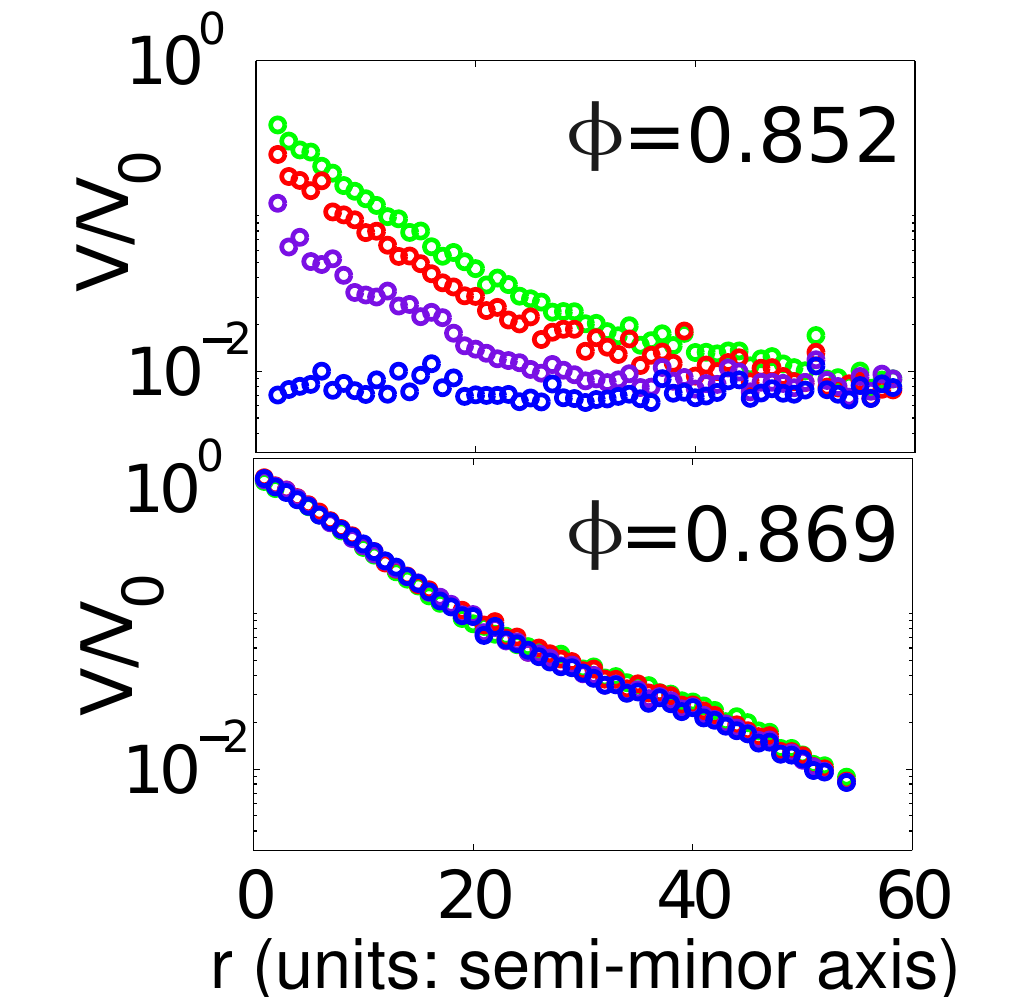}}
\caption{Schematics of the experimental setup. a) Side view. b) Small
  segments of polarized/unpolarized images. c) Top view of Couette
  cell. The inner wheel is lined with small teeth to increase the
  friction between the wheel and the particles.  Circularly polarized
  light passes through the particles from below, and a beam splitter
  above the experiment to produce two views of the same region of the
  experiment, each imaged by a different camera, with and without a
  (crossed) circular polarizer (relative to the original light
  polarization). Camera 1 records the particle photoelastic response,
  providing local stress information of the system\cite{sup1}; camera 2 yields a
  direct image of the particles\cite{sup2}. d) Radial profile of time-average
  velocities for two representative systems of ellipses. Colors
  represent the time interval (green:0-5, red:5-10, purple:10-15, and
  blue:15-20 revolutions).  $\Omega = 0.01$ rpm for all data.}
\label{fig:Couette_schematics}
\end{figure}

{\em Experiment:} The experimental data of this study were collected
from a quasi-2D Couette experiment\cite{howell99}. This geometry
allows continuous shear for arbitrary time/strain intervals. Using
this feature, we performed shearing experiments for up to 20
revolutions of the inner wheel, and collected two types of data. In
one case, we kept the shear rate constant at $\Omega =0.01$rpm, and
varied the global packing fraction, $\phi$. In the second case,
focused on rate dependence, we varied the $\Omega$ over a bit more
than a decade, for representative packing fractions. All the packing fractions were below $\phi_J$, which is $0.84$ for disks, and $0.91$ for ellipses \cite{farhadi-thesis12}.
Fig.\ref{fig:Couette_schematics} gives schematics of the experimental
setup. The particles, rest on a smooth horizontal Plexiglas sheet, and
are confined between an inner wheel and a rigid, concentric, outer
ring.  They are continuously sheared by slowly rotating the inner
wheel at constant rates, spanning $0.008 \leq \Omega \leq 0.128 rpm$.
Both ellipses and disks are made from the same photoelastic material
(Vishay polymer PSM-4), and are machined with similar
procedures. Thus, the particle surfaces have similar friction
coefficients. The ellipses have semi-minor axis $b\simeq0.25$cm and
aspect ratio $\sim2$. The radius of the small/large disks is
$r_s\simeq0.38$ cm/$r_l\simeq0.44$ cm. The number ratio of small-to-large
disks is kept at $\sim 9:2$. The radii of the inner and outer rings of
the Couette apparatus are $10.5$ cm and $25$ cm respectively. As the
system was sheared, two synchronized cameras obtained polarized and
unpolarized digital images every$~\sim10s$.  We studied systems of
either identical ellipses or bi-disperse collections of disks, for
packing fractions near the shear-jammed regime.  The unpolarized
images yield the centers, and orientations (for ellipses) for each
particle; the polarized images provide the local $P$ at the scale of a
particle using a measure, $g^{2}$, which is the average gradient
square of the photoelastic image intensity, integrated over a
particle; $g^{2}$ is proportional to local
pressure\cite{howell99,geng03}.  We used the green channel of our
images to compute $g^{2}$, since the photoelastic response is
color-dependent, and the polarizers are optimized for green.

We focus on several key results: 1) Anomalously slow relaxation for
the motion of ellipses at densities just below the shear-jamming
threshold; 2) Differences in the density profiles of disks and
ellipses; 3) Slow evolution of the average orientation of ellipses;
and 4) Strong, and previously unreported (to our knowledge) rate
dependence for stress fluctuations with ellipses that has not
counterpart for disks.  The first three of these are clearly linked.

\begin{figure}
\includegraphics[scale=0.35]{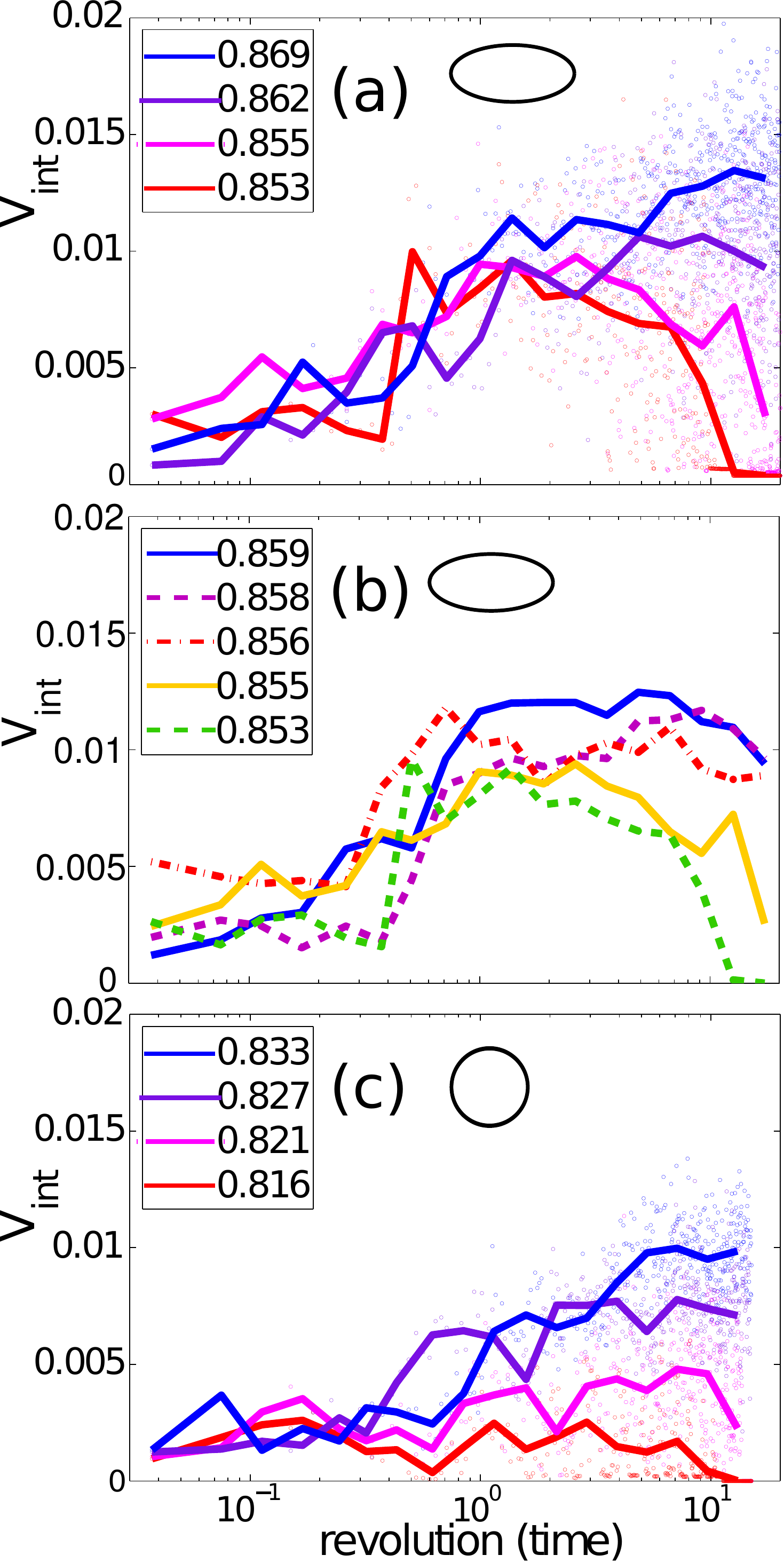}
\caption{ Evolution of $V_{int}$ for systems of a)ellipses, b)ellipses with $\phi$'s corresponding to meta-stable range, c)disks. The legends show global packing fractions. Data points demonstrate $V_{int}$, sampled from equal time intervals. The lines show the trend of mean value of $V_{int}$, averaged over logarithmically equal time intervals. }
\label{fig:velocity_drop}
\end{figure}

{\em Velocities:} We show the evolution of velocity profiles for ellipses at two typical $\phi$'s in
Fig.\ref{fig:Couette_schematics}(d). The velocity profile is stable
for denser states, but for slightly lower $\phi$, it relaxes slowly to
zero. To quantify the evolution of flow profile, we use
the average speed, $V_{int} = \frac{1}{A}\int <V> r dr/\int r dr$,
where, $V$ is the scalar value of the particle velocity
vector,$V=|\overrightarrow{v}|$, and $A$ is the area of the Couette cell.  Here, we first focus on a small range of $\phi$ where transient
behavior occurs for ellipses but not for disks.
Fig.\ref{fig:velocity_drop} shows the evolution of $V_{int}$ for ellipses and disks.
For ellipses, within a small but non-zero range of `meta-stable'
$\phi$'s, $V_{int}$ rises relatively quickly during the first
revolution to $V_{int}\simeq 0.01$ for all packing fractions, but then
drops slowly to zero.  For disks, we found no such meta-stable regime:
$V_{int}$ has an almost monotonic trend in the logarithmic time of an
experiment.  We show the changes in the meta-stable state region in a
narrow range of densities, $0.85\lesssim\phi\lesssim 0.86$, in
Fig.~\ref{fig:velocity_drop}b.  In this, and other figures throughout
the paper, systematic errors are at or below the scale of dots used to
indicate data points.  The apparent noise on the data is due to
statistical fluctuations, not to measurement errors.

\begin{figure}
\centering
\includegraphics[scale=0.47]{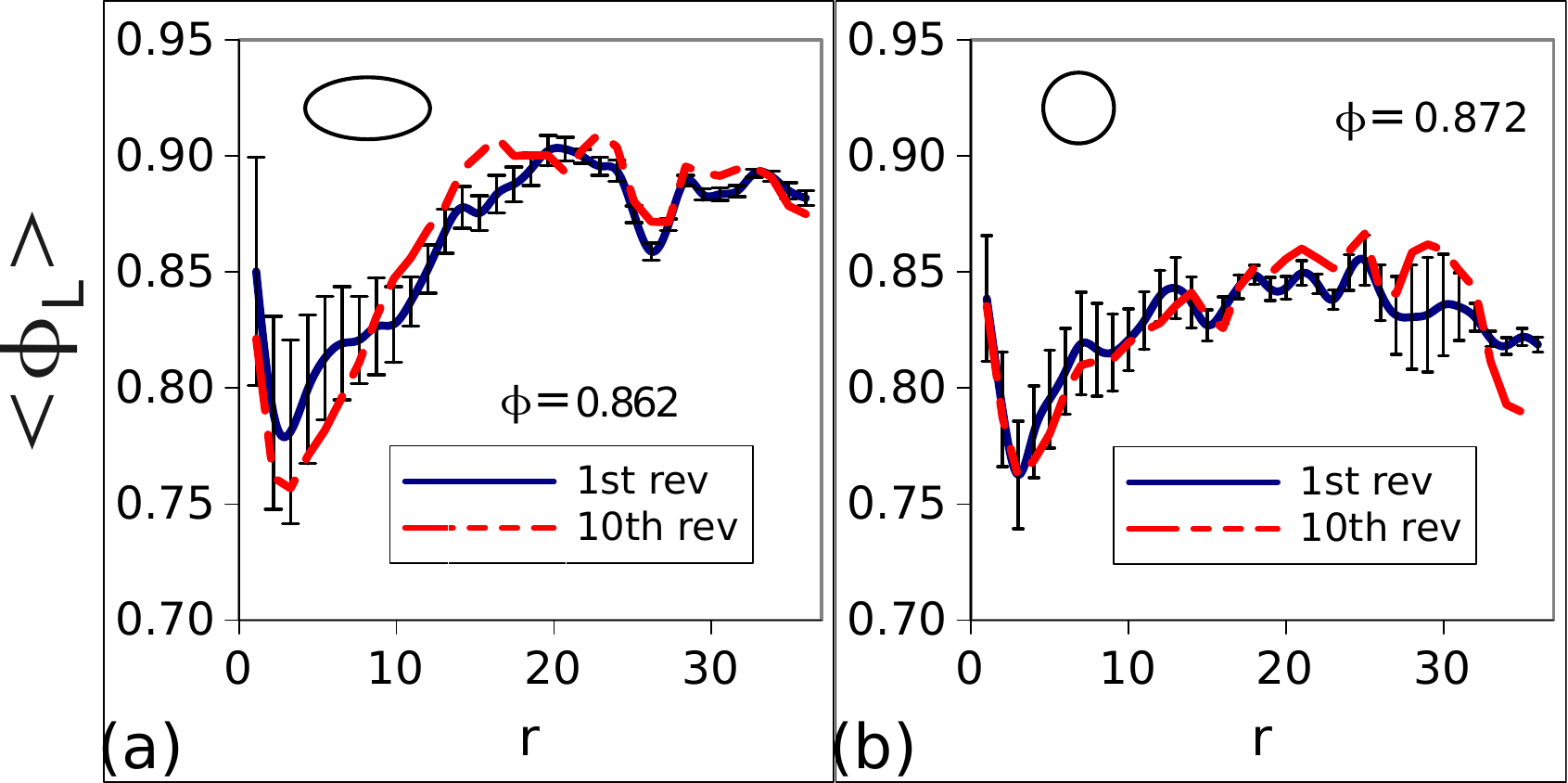}
\caption{Radial profiles of time-averaged local
  densities for ellipses, and disks. r (units: radius of large disk), is the radial
  distance from the inner wheel's perimeter.
  }
\label{fig:density_ellipses_disks}
\end{figure}

\begin{figure}
\centering
\includegraphics[scale=0.38]{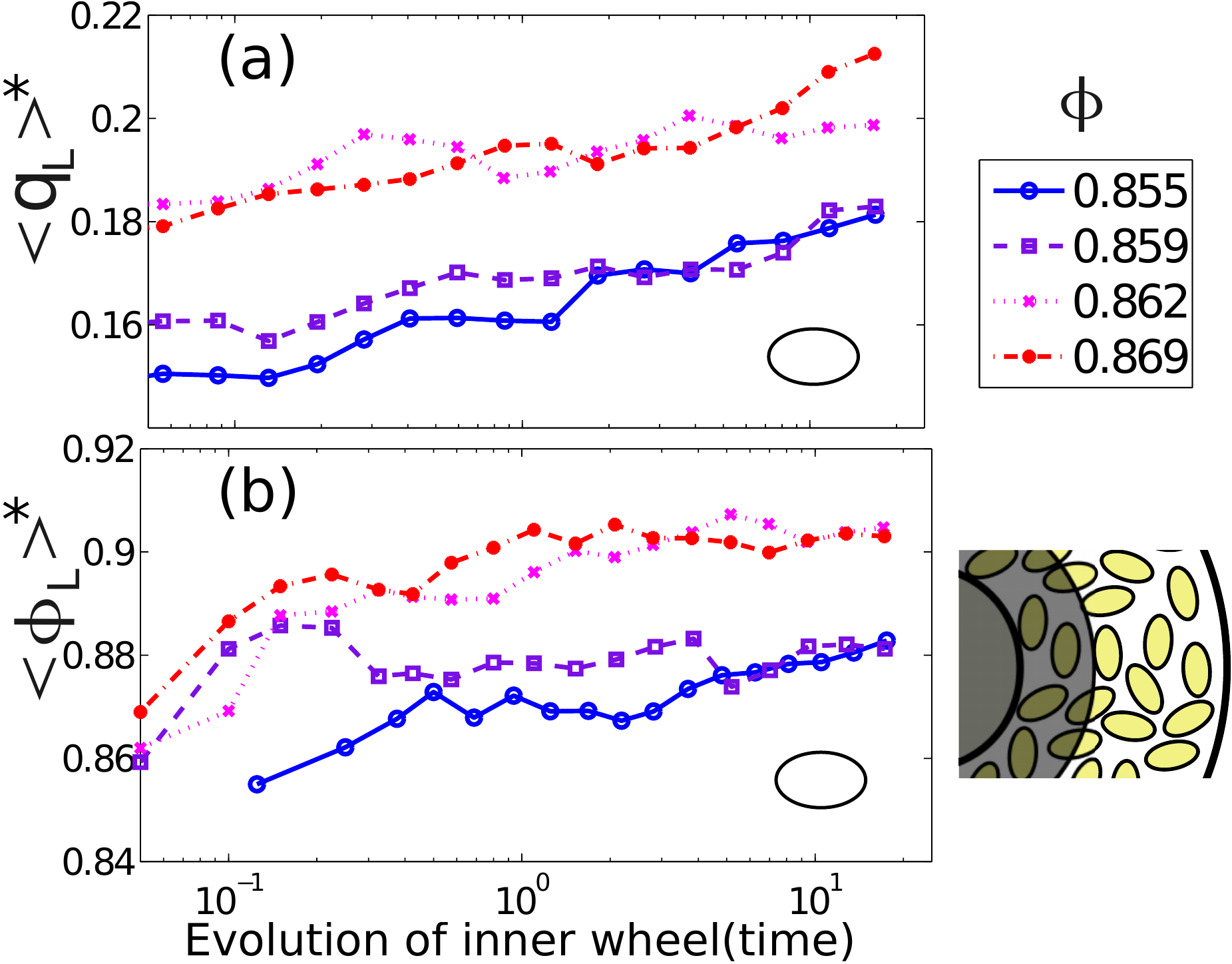}
\caption{Evolution of $<q_L>^*$ and $<\phi_L>^*$ for various packing
  fractions (legend). $<q>^*$ represents the average local
  orientational order parameter for the area shown as unshaded in
  schematics shown in lower right, The excluded region is about
  $16\times$ semi-minor from the inner wheel surface, i.e. roughly the
  radial width of the shear band.  }
\label{fig:order_par_global}
\end{figure}

\begin{figure}
\includegraphics[scale=0.317]{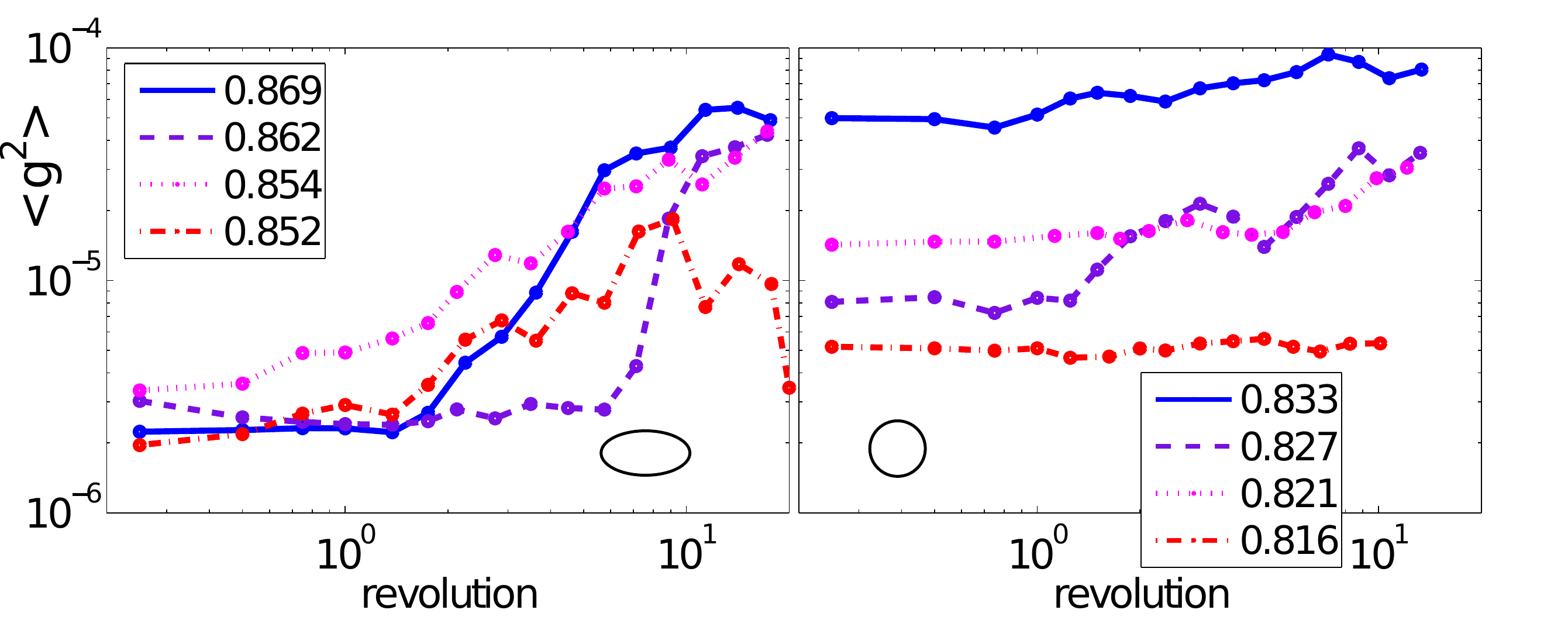}
\caption{Evolution of  $<g^2>$ for a) ellipses, and b) disks. Legends value indicate the global packing fraction for each data set.}
\label{fig:pressure_evolution}
\end{figure}

{\em Structure: local density and orientational order:} The gradual
decrease in $V_{int}$ for the meta-stable regime of ellipses is
associated with a gradual, and anomalously large, increase in Reynolds
dilatancy in the region close to the inner wheel, which is facilitated
by the evolution of the ellipse orientation.  We measure the local
density by determining the Voronoi areas, $V^v$, for individual
particles\cite{farhadi-thesis12}.  
We take the
  Voronoi area for a given particle to correspond to the interior area
  of the particle, plus the area corresponding to all points in the
  plane that are closer to the boundary of the given particle than to
  the boundary of any other particle.
    The local solid fraction in
the region enclosed by $V^v$ is $\phi _L= \frac{V_0}{V^v}$, where
$V_0$ is the area of a particle. Fig.~\ref{fig:density_ellipses_disks}
shows radially and time averaged local density profiles extracted from Voronoi analysis. 
Close to the inner wheel,
  both types of particles show a dip in density, i.e. a shear band.
 Also, the drop in density
  is roughly twice as big for ellipses (e.g. as a fraction of the mean
  packing fraction) as for the disks. 
   The small rise in $\phi$ at
the inner wheel is due to
the carrying of particles by inner shearing wheel.
 Since the global packing fraction is fixed
for each data set, the relative density minimum near the inner wheel
induces closer packing in the middle and outer radial regions of the
Couette cell. For all measurements of structure, e.g. for
Figs.~\ref{fig:velocity_drop}-~\ref{fig:pressure_evolution}, we keep
the rotation rate of the inner wheel fixed at $\Omega=0.01$rpm.  
At the end of this paper, we consider the effects of varying the rotation rate.

It is known that random packing of ellipses can be denser than random
disks
\cite{donev04}. However, compacting a randomly prepared
system of ellipses requires reorientation of the particles which leads
to locally nematic order, as characterized by the local mean
\emph{director}.  The director has a similar sense as for the liquid
crystal context. In order to quantify nematic ordering, we determine
the local order parameter $q_L$. 
Here, 
$q_L$, is computed from the
orientational order matrix, $Q$, calculated for small patches of $N
\simeq 10$ ellipses surrounding the i'th particle: $Q_{xy}=
\frac{1}{N}\sum_{n} (u_x^{n}u_{y}^{n} - \frac{1}{2} \delta^{xy})$
\cite{chaikin95}.  The size of a patch was chosen to give a local
average.  Here, $x$ and $y$ are two given orthogonal directions, and
$u^n$ is the unit vector in the direction of the major axis for a
given particle. The summation is made over all particles in the
patch. We determine $q_L$ as the maximum eigenvalue of $Q$.

Fig~\ref{fig:order_par_global}a shows the evolution of $<q_L>^{*}$,
the spatially average of $q_L$, for the outer region of the cell
vs. rotations of the inner wheel (on a logarithmic scale).
Specifically, we average $q_L$ over a radial region $16 a \leq r \leq
R_{o}$, where $a$ is the semi-minor axis of the ellipses and $R_o$ is
the radius of the outer confining ring of the Couette cell.  In this
region, the average velocity is about $10\%$ of the average velocity
next to the inner wheel. Considering only this region enables us to
characterize the orientational ordering of particles which are almost
stationary, but at the same time are reoriented gradually by the
dynamics generated in the shear band.
In Fig~\ref{fig:order_par_global}a, for all
densities, including stable and meta-stable values, $<q_L>^{*}$
increases gradually in time, roughly as $log(t)$. The evolution of
average local densities (from Voronoi analysis) in the same outer
region, $<\phi_L>^{*}$, is shown in Fig~\ref{fig:order_par_global}b.
Both density and orientational order increase with time, although the
relation between these two may not be strictly linear.  This extremely
slow reorientation and compaction is driven by/allows more dilation in
the inner layers.  It is then the source of the meta-stability and it
subsequently halts the shearing when the compaction in the outer
region leads to too weak mechanical contact between the shearing wheel
and the ellipses.

{\em Evolution of average pressure:}
Fig.~\ref{fig:pressure_evolution} shows the evolution of the
system-average of $g^2$ for representative packing fractions of disks
and ellipses. As mentioned, $g^2$ quantifies global pressure. For disks, $<g^2>$ attains its initial
value relatively quickly in most cases, except for a slight increase over the course of an
experiment. However, for ellipses, $<g^2>$ steadily increases over a
span of about 10 revolutions of the inner wheel. After this long
initial increase, the average $g^2$ drops only for meta-stable systems
and saturates for densities above this range.

\begin{figure}
\centering
\includegraphics[scale=0.35]{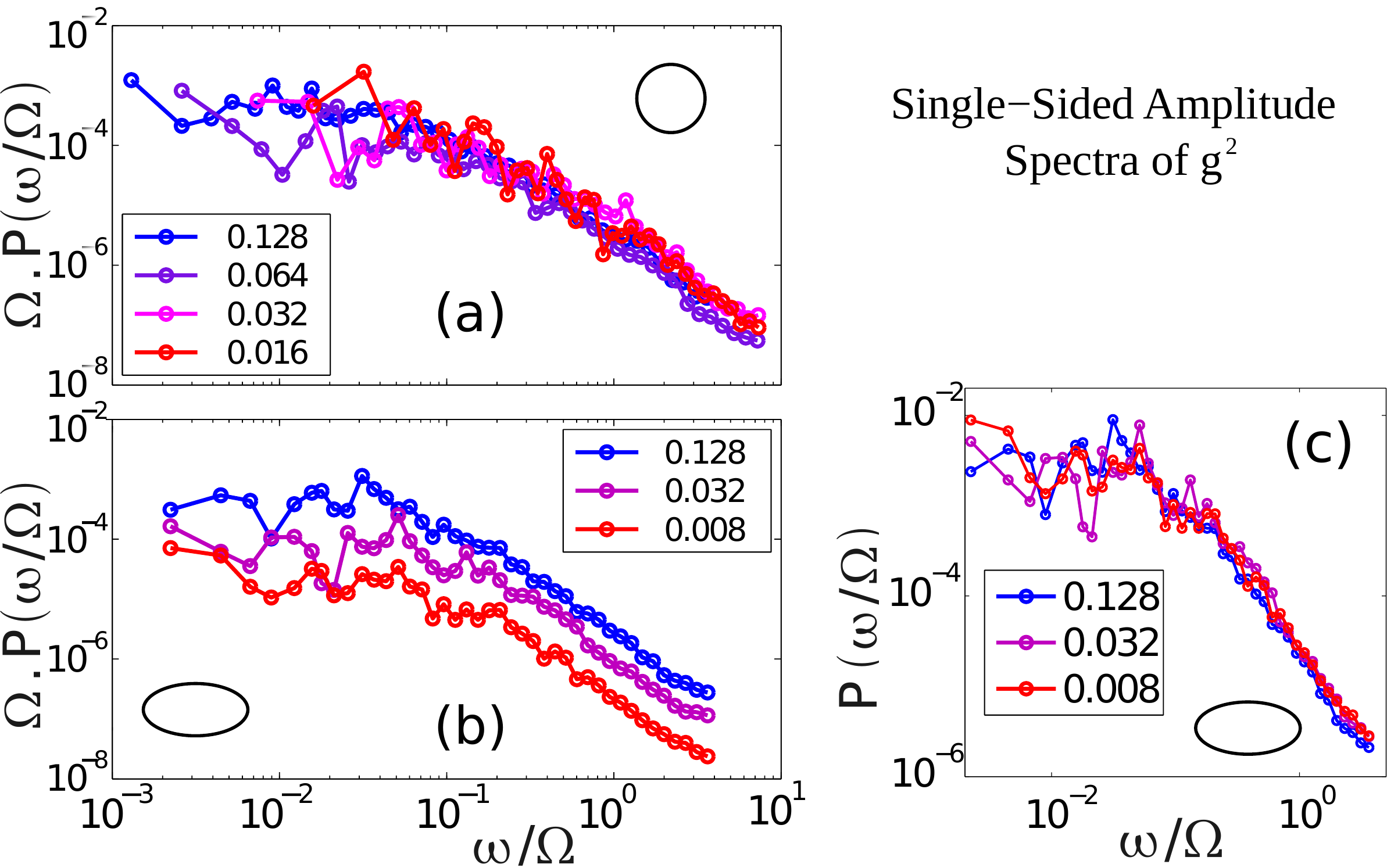}
\caption{Scaled power spectral density (PSD) of $g^{2}$ for different
  shear rates, $\Omega$, for systems of: a) disks with $\phi=0.83$,
  and b) ellipses with $\phi=0.86$. c) Here, vertical axis of part b
  (ellipses) is rescaled by $\Omega$. Note that these data are taken
  for packing fractions in the stable region, where the velocity
  profile persists for very large strains (see
  Fig.\ref{fig:Couette_schematics}d) The legends indicate the value of
  $\Omega$ in rpm for each data set.
}
\label{fig:spectra_collapse}
\end{figure}

{\em Rate dependence of pressure fluctuations:} Perhaps the most
significant difference in rate effects for ellipses vs. disks occurs
for time fluctuations of the pressure in the stable
regime. Specifically, the instantaneous pressure of the system is
proportional to $g^2$ integrated over the whole field of view.  Here,
we consider the fluctuations, $\delta g^2(t)$, in time-series of this global
$g^2(t)$. Up to this point, the shear rate was held fixed at
$\Omega=0.01$rpm.  Now, we consider the effect of changing $\Omega$,
by allowing this quantity to vary over $0.008 \leq \Omega \leq
0.128$rpm.  

The concept of rate-independence is based on the fact that if the strains
are slow enough, the system remains very close to a sequence of states
in mechanical equilibrium.  Then, the system response, for instance the pressure, would depend on the strain, but not the rate of strain, $\dot{\gamma}$.
By contrast, if 
$\dot{\gamma}$
were too high, the system would be out of mechanical equilibrium, and
the response would explicitly depend on $\dot{\gamma}$ (here, $\Omega$).  The power spectrum,
$P(\omega)$, for fluctuations of $g^2(t)$ provide a simple test of
whether the system is rate independent or not.
Normally, $P(\omega) = T^{-1} |\int_0^T \delta g^2(t) \exp(-i\omega t) dt|^2$, where $T$ is the time over which $\delta g^2(t)=g^2(t)-<g^2>_t$ is
  measured, and $\omega$ is the spectral frequency.  If the fluctuations were statistically rate independent,
  then it would be possible to replace time, $t$, with strain, $\gamma
  = \dot{\gamma} t$, or here, $\Omega$.  We would then find the same
  power spectra regardless of $\Omega$ for spectra computed with
  strain as the variable, instead of time, as discussed for instance
  by Miller et al.\cite{miller96}.  A straight forward change of
  variables from $t$ to $\gamma = \dot{\gamma} t$ shows that for the
  rate independent case, $\Omega P(\omega / \Omega)$ is a function
  only of $\omega/\Omega$. Fig.\ref{fig:spectra_collapse}a,b show the scaled power
  spectra computed from $g^2(t)$ for various
  $\Omega$, where for all cases, the data pertain to stable
  states. The spectra collapse reasonably well to one generic curve in
  systems of bi-disperse disks.  Hence, the fluctuations of global
  stress for disks are nearly invariant under changes of shear rate
  over the range of $\Omega$ studied here. (Note that this does not
  exclude the possibility of weak, e.g. logarithmic rate effects.)
  These results for disks are qualitatively similar to
    observations by Miller et al.\cite{miller96} for fluctuations for
    systems of sheared 3D spheres.  By contrast, similar data for
    systems of ellipses (Fig.\ref{fig:spectra_collapse}b) show no
    collapse, which implies a strong rate dependence. In particular,
    we observe that the shift in scaled spectra is proportional to the
    shear rate $\Omega$, Fig.\ref{fig:spectra_collapse}c. The data
    suggest that the {\em fluctuations} in pressure, $\delta g^2$, have
    a form that is roughly $\delta g^2(t) = \dot{\gamma}^{1/2}
    f(\dot{\gamma} t)$. This means $\delta g^2$ consists of a rate
    independent term ($f(\dot{\gamma} t)$), times the square root of the shear rate.  That
    is, there is explicit non-trivial rate dependence, which is neither
    viscous, nor quasi-static. We note that this is a dynamical effect, and to our knowledge, it has not been previously reported.
  


{\em Conclusions:} We have shown through parallel studies using
elliptical and circular quasi-2D particles that the long-term
rheological behavior of 2D shear jammed granular systems depends
significantly on particle shape. The difference in response of
ellipses and disks takes several forms, which depend on density.  One
difference involves transient dynamics in a narrow range of densities,
where ellipses show `meta-stable' dynamics that eventually lead to
zero-stress states.  The velocity, e.g. $V_{int}$, and the pressure
grow to a maxima, and then decay slowly to $0$, over large strains, as
the particles reorient and compact in the nearly static regions.  For
densities above the meta-stable regime, the radial profile of $\phi$
shows a significant minimum for ellipses.  The density dip in the
shear band is weaker for systems of disks, which also do not exhibit
meta-stable states.  The slow relaxation of velocity profiles and
local densities for ellipses are coupled to logarithmically slow
evolution in the local orientational order parameter of ellipses in
the compacted region far from the shear band.  This reorientation of
particles incrementally increases the packing fraction far from the
inner wheel, and allows the system to dilate more in the inner region
where the shear band is located.  The anomalous rate dependence seen
for ellipses is therefore tied to the orientational degree of freedom.
Although logarithmic rate dependence has been observed for
spherical/circular particles, ellipses show much stronger rate
dependence for stress fluctuations.  

The present experiments suggest that, more generally, when the density
is coupled to another system property, there exists the possibility of
a long internal response time as the system responds to shear or other
strains by forming more compact structures.  We note that complex
grain shapes do not have any obvious orientational order.  But, they
do have the property that rotations of the grains can lead to changes
of density or stresses.  These changes might be either positive or
negative, depending on the initial conditions and the strain history.

We thank Dr. Joshua Dijksman  for helping with setting up these
experiments, and we very much appreciate discussions with Prof. Karin
Dahmen.  This work was supported by NSF grants DMR-0906908,
DMR-1206351, and ARO grant W911NF-11-1-0110.

\end{document}